\pdfoutput=1
\documentclass[11pt]{article}

\usepackage[margin=1in]{geometry}
\usepackage{mathptmx}
\usepackage[T1]{fontenc}
\usepackage{graphicx}
\usepackage{booktabs}
\usepackage{amsmath}
\usepackage{array}
\usepackage{tabularx}
\usepackage{ragged2e}
\usepackage{caption}
\usepackage[activate={true,nocompatibility},final,tracking=false,kerning=false,spacing=false,factor=1100]{microtype}
\microtypesetup{expansion=false}
\usepackage[numbers,sort&compress]{natbib}
\usepackage[hidelinks]{hyperref}
\usepackage{xcolor}

\newcolumntype{L}[1]{>{\RaggedRight\arraybackslash}p{#1}}

\title{\textbf{The Multi-Lab Enterprise:\\ Governance, FinOps, and Telemetry Challenges\\ of Multi-Model AI Adoption}}

\author{
  Fabricio F. Costa, PhD, MBA, PMP$^{1,2,3,4,*}$\\[0.8em]
  \small $^{1}$\textit{AIx4All, LLC, Sunnyvale, CA 94087, USA}\\
  \small $^{2}$\textit{HCLTech, 2600 Great America Way, Suite 101, Santa Clara, CA 95054, USA}\\
  \small $^{3}$\textit{Genomic Sciences and Biotechnology Program, UCB, Bras\'ilia, DF 70.790-160, Brazil}\\
  \small $^{4}$\textit{Cancer Biology and Epigenomics Program, Stanley Manne Children's Research Institute,}\\
  \small \textit{Ann \& Robert H. Lurie Children's Hospital of Chicago, Northwestern University}\\
  \small \textit{Feinberg School of Medicine, Chicago, IL 60614, USA}\\[0.5em]
  \small $^{*}$Corresponding author: \texttt{fcosta@aix4all.com}; \texttt{fabriciof.costa@hotmail.com}
}

\date{\small This version: 2026. Preprint --- comments welcome.}

\begin{document}

\maketitle

\begin{abstract}
\noindent Enterprises are not choosing a single frontier AI provider; they are licensing all of them. As of early 2026, 81\% of Global~2000 enterprises run three or more model families, and OpenAI, Anthropic, and Google Gemini together account for roughly 88 to 89\% of enterprise LLM usage and spend. Drawing on survey data, transaction data, provider disclosures, and case studies across finance, legal, consulting, healthcare, life sciences, retail, and government, this paper shows that multi-lab licensing is a structural feature of the market, driven by durable task-specific model differentiation rather than a transitional phase awaiting commoditization. This structure creates three operational problems. Governance fragmentation (P1): heterogeneous vendor security postures, documentation, and compliance surfaces must be reconciled across overlapping regulatory frameworks while shadow AI proliferates. FinOps breakdown (P2): token-based, behavior-driven consumption defeats budgeting. In the past year, 79\% of enterprises overran AI budgets, with FinOps-mature organizations overshooting by a mean of 30.9\%, and no standardized cross-provider unit of spend exists. Telemetry fragmentation (P3): each lab exposes adoption and cost data through incompatible consoles, APIs, and metric definitions, forcing bespoke unification layers. We map the emerging responses, including LLM gateways, observability platforms, and the Tokenomics Foundation's FOCUS extension. We then develop a five-metric framework for evaluating API and agent cost burn, with a worked example where the cheapest model per attempt is the most expensive per successful task. We conclude that P1, P2, and P3 reflect one missing abstraction: a cross-provider enterprise AI control plane.
\end{abstract}

\begin{center}
\small\textit{The views expressed are the author's own.}
\end{center}

\vspace{0.5em}

\section{Introduction}
\label{sec:intro}

In less than three years, generative AI has become the fastest-scaling software category in history: enterprise spending on generative AI grew from \$1.7B in 2023 to \$11.5B in 2024 to \$37B in 2025~\cite{menlo2025}. Beneath this growth curve, however, lies a market structure that contradicts the standard enterprise-software playbook. Enterprises do not standardize on one vendor. They license OpenAI \emph{and} Anthropic \emph{and} Google---often simultaneously, often within the same quarter, and sometimes for the same workload class.

The evidence is consistent across independent data sources. A January 2026 survey of 100 Global~2000 executives found 81\% of enterprises using three or more model families, up from 68\% a year earlier~\cite{a16z2026}. Transaction data across more than 70,000 U.S.\ firms shows the most sophisticated AI buyers deliberately avoiding single-vendor concentration~\cite{ramp2026}. Even the U.S.\ federal government, through the General Services Administration's OneGov agreements of August 2025, licensed ChatGPT Enterprise, Claude for Government, and Gemini for Government within a fifteen-day window---at nominal fees of \$1, \$1, and \$0.47 per agency respectively---explicitly enabling agencies to adopt any or all~\cite{gsa2025onegov}.

This paper asks two questions:

\begin{itemize}
  \item \textbf{RQ1:} Why do enterprises across verticals license all top-tier frontier AI labs rather than standardizing on one?
  \item \textbf{RQ2:} What operational problems does this multi-lab structure create, and how are enterprises and the surrounding ecosystem responding?
\end{itemize}

We argue that the answer to RQ1 is \emph{durable differentiation}: frontier models have diverged, not converged, in task-specific capability---coding and agentic work, long-context multimodal processing, general knowledge work and ecosystem breadth---and enterprise switching behavior confirms that buyers treat labs as complements rather than substitutes. The answer to RQ2 is organized around three problems, which we label following the numbering of the practitioner observations that motivated this work:

\begin{description}
  \item[P1 --- Governance fragmentation.] Each additional lab multiplies the compliance surface: separate security reviews, heterogeneous model documentation, distinct data-processing agreements, and divergent certification postures, all mapped against overlapping regulatory frameworks (NIST AI RMF, ISO/IEC 42001, the EU AI Act, and sector-specific regimes).
  \item[P2 --- FinOps breakdown.] The atomic unit of AI cost is the token, whose consumption scales with model behavior---prompt length, reasoning depth, agentic iteration---rather than with seats or headcount. Balancing token consumption against dollar budgets across providers with incompatible pricing structures defeats traditional financial planning.
  \item[P3 --- Telemetry fragmentation.] Each lab exposes adoption and spend data through its own administrative console and APIs, with incompatible metric definitions, export formats, and refresh latencies. No enterprise can answer ``how much are we spending on AI, on what, and to what effect?'' from any single vendor surface.
\end{description}

\begin{figure}[t]
  \centering
  \includegraphics[width=0.9\textwidth]{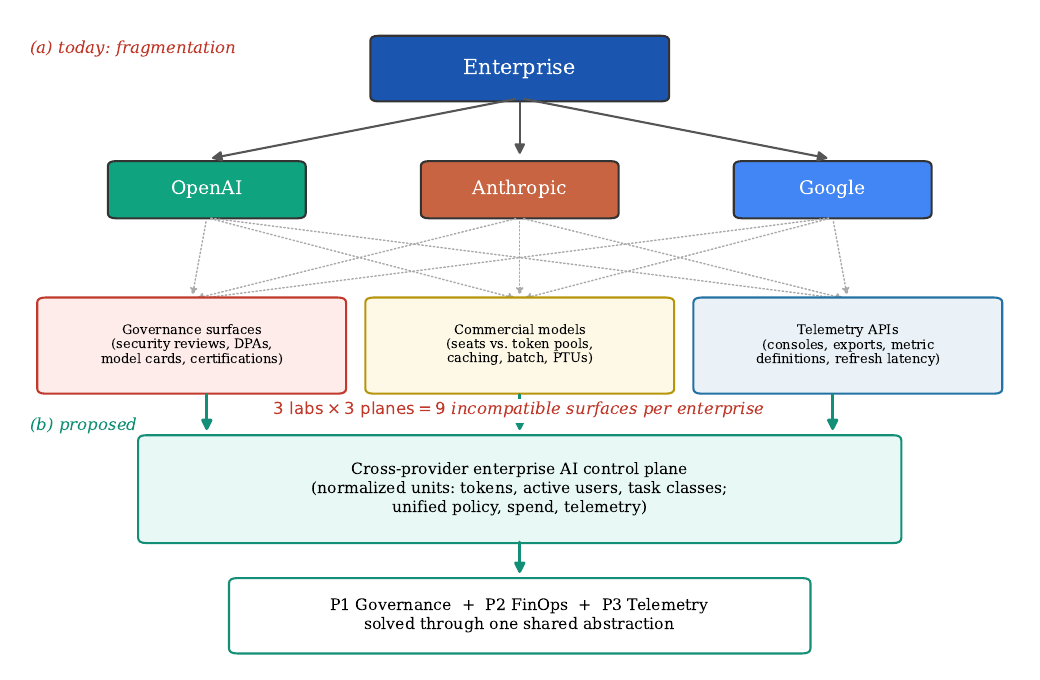}
  \caption{The paper's organizing framework. (a)~Today, each licensed lab exposes its own governance surface (P1), commercial model (P2), and telemetry API (P3), yielding $3\times3=9$ incompatible surfaces per three-lab enterprise. (b)~The proposed cross-provider control plane normalizes policy, spend, and usage into shared units, addressing P1--P3 through one abstraction (\S\ref{sec:discussion}).}
  \label{fig:controlplane}
\end{figure}

Figure~\ref{fig:controlplane} previews the paper's organizing argument: three labs multiplied by three incompatible operational planes yields nine surfaces every multi-lab enterprise must reconcile alone, and P1--P3 dissolve jointly---or not at all---through a shared control-plane abstraction developed in Section~\ref{sec:discussion}.

\paragraph{Contributions.} This paper makes four contributions. (1)~A triangulated empirical account of enterprise multi-lab licensing that combines survey, transaction, and disclosure data with explicit treatment of source conflicts of interest. (2)~A three-problem framework (P1/P2/P3) grounded in documented enterprise cases across more than seven verticals. (3)~A vendor-neutral structured comparison of provider administrative and usage-telemetry surfaces (Table~\ref{tab:telemetry}), which to our knowledge is the first of its kind. (4)~A minimal five-metric framework for evaluating API and agent cost burn (cost per attempt, cost-of-pass, burn multiplier, difficulty elasticity, marginal cost of capability), with a worked example. (5)~A gap analysis positioning the technical LLM-routing literature against the organizational problems enterprises face, together with a research and standardization agenda.

\section{Background and Related Work}
\label{sec:background}

\subsection{Enterprise LLM Market Structure, 2023--2026}

The enterprise LLM market has undergone a rapid redistribution of share. Anthropic's share of enterprise LLM spend grew from 12\% in 2023 to 24\% in 2024 to 40\% in 2025, while OpenAI's declined from 50\% to 27\% and Google's rose from 7\% to 21\% over the same period~\cite{menlo2025,menlomid2025} (Figure~\ref{fig:marketshare}). Total enterprise generative AI spending reached \$37B in 2025, with the application layer accounting for \$19B~\cite{menlo2025} (Figure~\ref{fig:marketshare}b). Average per-enterprise LLM spend rose from approximately \$4.5M to \$7M over two years and is projected to reach \$11.6M~\cite{a16z2026}.

Critically for this paper's thesis, provider-level stickiness coexists with portfolio-level expansion: only 11\% of enterprise teams changed model providers in the past year, while 66\% upgraded to a newer model from an existing vendor~\cite{menlo2025}. Enterprises rarely \emph{substitute} labs; they \emph{add} them.

\subsection{The LLM Routing and Orchestration Literature}

A substantial technical literature addresses cost-quality optimization across multiple LLMs. FrugalGPT demonstrated cascade strategies reducing inference cost by up to 98\% while matching top-model quality~\cite{chen2023frugalgpt}. RouteLLM learns routing policies from preference data, achieving cost reductions of 85\% on MT-Bench and 45\% on MMLU~\cite{ong2025routellm}. Related systems include Hybrid LLM~\cite{ding2024hybrid}, RouterDC~\cite{chen2024routerdc}, MoDEM~\cite{simonds2024modem}, MetaLLM~\cite{nguyen2024metallm}, and GraphRouter~\cite{feng2024graphrouter}, with RouterBench~\cite{hu2024routerbench} and RouterArena~\cite{lu2025routerarena} providing standardized evaluation.

This literature answers a narrower question than the one enterprises face. Routing optimizes \emph{which model answers a given query}. It does not address who reviews four vendors' security postures (P1), how finance reconciles four incompatible billing models (P2), or how adoption is measured consistently across four administrative consoles (P3). The organizational layer above routing is, to our knowledge, unaddressed in the peer-reviewed literature; the present paper targets that gap.

\subsection{AI Governance Frameworks}

Enterprises deploying frontier models operate under a growing stack of overlapping frameworks: the voluntary NIST AI Risk Management Framework~\cite{nistairmf}; the certifiable ISO/IEC 42001 AI management-system standard~\cite{iso42001}, which industry surveys suggest a large majority of organizations plan to pursue; and the mandatory EU AI Act~\cite{euaiact}, whose obligations phase in between February 2025 and August 2027. Sector regimes add further constraints: HIPAA and FDA guidance in healthcare and life sciences, and FINRA/SEC/OCC expectations in financial services. Section~\ref{sec:p1} analyzes how multi-lab licensing multiplies the burden of complying with this stack.

\subsection{FinOps and the Token as a Cost Unit}

The FinOps Foundation's FOCUS specification standardized cloud cost-and-usage data, but tokens---the atomic unit of LLM consumption---fell outside its original scope. In 2026 the Linux Foundation announced the Tokenomics Foundation, in collaboration with the FinOps Foundation, to extend FOCUS to token-based consumption; initial supporters include Accenture, Google Cloud, IBM, JPMorganChase, KPMG, Microsoft, Oracle, Salesforce, SAP, and ServiceNow~\cite{tokenomics2026}. The State of FinOps 2026 survey reports that AI cost management became a near-universal practice concern, cited by 98\% of respondents, up from 63\% a year earlier~\cite{finops2026}. Section~\ref{sec:p2} builds on this backdrop.

\section{Data and Methods}
\label{sec:methods}

This is a synthesis study. We combine four evidence classes with distinct reliability properties, and we treat their disagreements as informative rather than as noise to be averaged away.

\begin{enumerate}
  \item \textbf{Survey data:} Menlo Ventures (n${\approx}$500 U.S.\ enterprise decision-makers, November 2025; n=150 technical leaders, July 2025)~\cite{menlo2025,menlomid2025}; Andreessen Horowitz (n=100 CIOs, 2025; n=100 Global~2000 executives, January 2026)~\cite{a16z2025,a16z2026}; McKinsey State of AI (2025)~\cite{mckinsey2025}; DoiT/Sapio (n=500 finance leaders, $\pm$4.4pp at 95\% CI)~\cite{doit2026}; Mavvrik/Benchmarkit (n=372 enterprises)~\cite{mavvrik2025}.
  \item \textbf{Transaction data:} the Ramp AI Index, derived from corporate card and bill-pay flows across more than 70,000 U.S.\ businesses~\cite{ramp2026}. Transaction data is behavioral rather than self-reported and uniquely captures decentralized ``shadow'' purchasing.
  \item \textbf{Provider disclosures:} earnings releases and executive statements from Microsoft, Alphabet, Amazon, OpenAI, and Anthropic~\cite{msft2026q3,alphabet2026q2,amazon2026q1,reuters2026anthropic}. These are flagged as company-claimed throughout.
  \item \textbf{Case documentation:} primary company announcements, GSA award notices, and credible trade press underpinning the vertical case matrix of Section~\ref{sec:vertical}.
\end{enumerate}

\begin{figure}[t]
  \centering
  \includegraphics[width=0.85\textwidth]{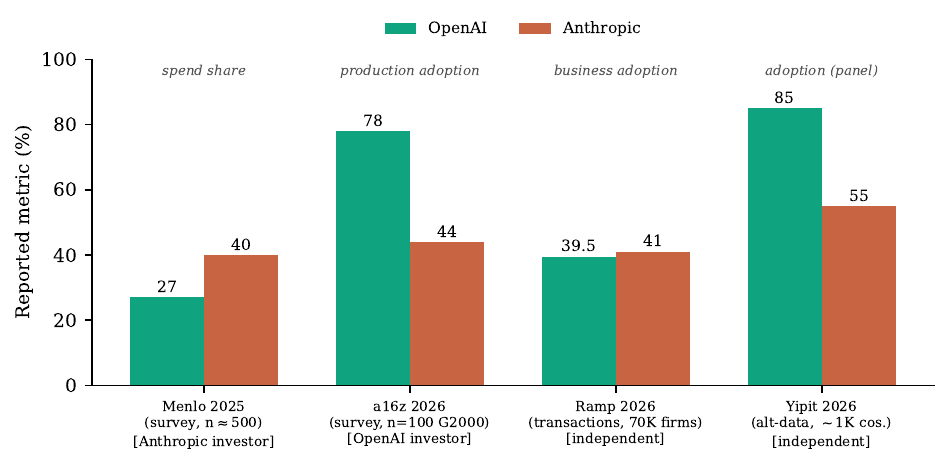}
  \caption{Source triangulation for OpenAI and Anthropic enterprise position across four data series with different constructs and different conflicts of interest~\cite{menlo2025,a16z2026,ramp2026}. Divergence across series is treated as information, not noise (\S\ref{sec:methods}).}
  \label{fig:triangulation}
\end{figure}

\begin{figure}[t]
  \centering
  \includegraphics[width=\textwidth]{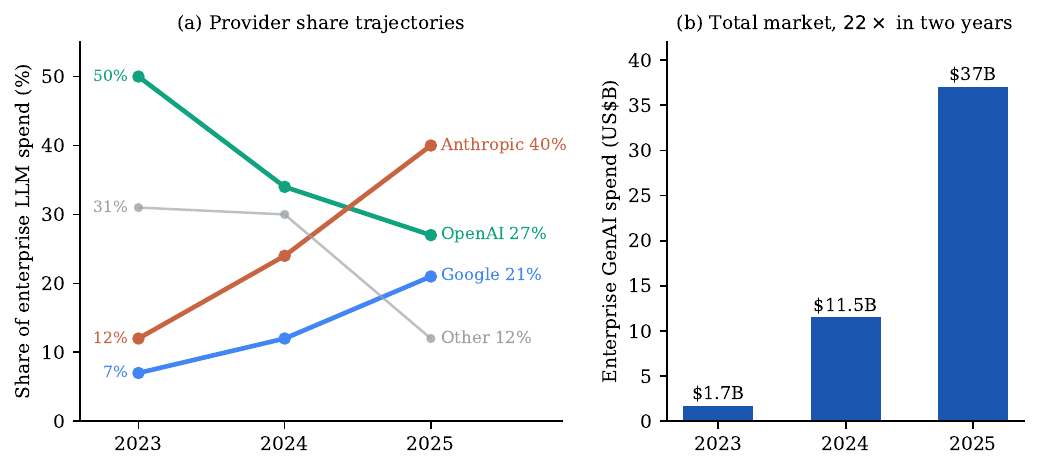}
  \caption{The redistribution, not contraction, of the enterprise LLM market. (a)~Provider share trajectories, 2023--2025: Anthropic and OpenAI cross as Anthropic climbs 12\%$\to$40\% and OpenAI declines 50\%$\to$27\%. (b)~The market itself grew 22$\times$ (\$1.7B$\to$\$37B), so every provider's absolute revenue grew even as shares shifted. Data: Menlo Ventures annual surveys~\cite{menlo2025,menlomid2025}; 2024 OpenAI value interpolated from Menlo's 2024 report. Menlo is an Anthropic investor (\S\ref{sec:methods}).}
  \label{fig:marketshare}
\end{figure}

\paragraph{Conflict-of-interest handling.} The two most-cited enterprise AI market surveys are published by venture firms with direct financial stakes in the outcome: Menlo Ventures is an Anthropic investor, and Andreessen Horowitz is an OpenAI investor. Anthropic additionally reports revenue transacted through cloud resellers on a gross basis, complicating cross-lab revenue comparison~\cite{sacra2026}. Our mitigation is triangulation: where survey, transaction, and disclosure series diverge, we report the range and note the direction of plausible bias rather than selecting a single point estimate. Where independent corroboration exists---for example, an independent alternative-data panel of roughly 1{,}000 companies finding OpenAI adoption near 85\% and Anthropic near 55\%, consistent with the a16z survey~\cite{a16z2026}---we say so. Figure~\ref{fig:triangulation} makes the triangulation explicit: the four series measure related but non-identical constructs (spend share, production adoption, paid business adoption, panel adoption), which explains part of their divergence; the residual divergence tracks the direction one would predict from each publisher's financial position, which is exactly why no single series should be cited as ground truth.

\paragraph{Limitations.} The quantitative base is U.S.-centric; survey sampling frames differ across sources and years; and the market moves quickly, so all figures are dated to their publication month. This paper characterizes a market circa mid-2026; its framework (P1--P3) is intended to outlive its point estimates.

\section{The Multi-Lab Licensing Phenomenon}
\label{sec:phenomenon}

\subsection{Prevalence}

\begin{figure}[t]
  \centering
  \includegraphics[width=0.9\textwidth]{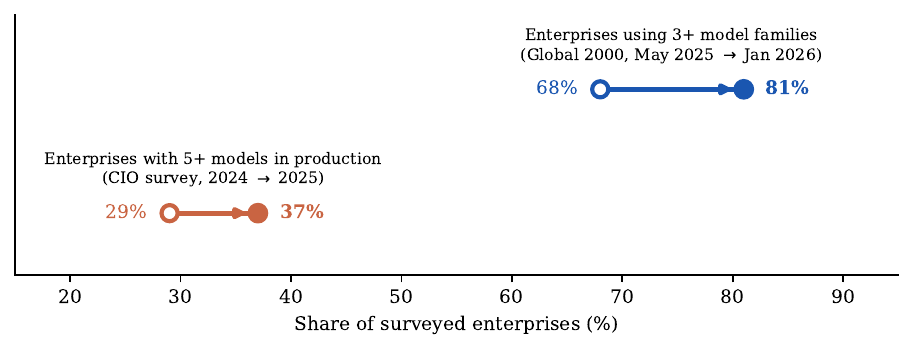}
  \caption{Multi-homing is deepening on both available measures: three-plus model families among Global~2000 enterprises (68\%$\to$81\%, May 2025$\to$Jan 2026; n=100)~\cite{a16z2026} and five-plus models in production (29\%$\to$37\%, 2024$\to$2025; n=100 CIOs)~\cite{a16z2025}. Arrows indicate direction of change between survey waves.}
  \label{fig:modelcounts}
\end{figure}

Multi-lab licensing is now the dominant enterprise pattern. As of January 2026, 81\% of surveyed Global~2000 enterprises used three or more model families in testing or production, up from 68\% in mid-2025~\cite{a16z2026}; 37\% of enterprises ran five or more models in production as of 2025, up from 29\% the year prior~\cite{a16z2025} (Figure~\ref{fig:modelcounts}). OpenAI models were in production at 78\% of surveyed enterprises; Anthropic reached 44\% in production and over 63\% including testing, the largest penetration gain of any lab over the survey interval~\cite{a16z2026}. The three leading labs jointly account for approximately 88--89\% of enterprise LLM spend and wallet share~\cite{menlo2025,a16z2026}.

\subsection{Why All of Them: Durable Differentiation by Task}
\label{sec:differentiation}

The central explanatory claim of this paper is that enterprises license all top-tier labs because the labs have \emph{diverged} in task-specific capability, making them complements rather than substitutes.

\textbf{Coding and agentic work.} AI-assisted software development is a category of roughly \$4B in enterprise spend, within which Anthropic holds a 54\% share~\cite{menlo2025}; at mid-2025, Claude models held 42\% of developer usage against OpenAI's 21\%~\cite{menlomid2025}.

\textbf{Reasoning.} 54\% of surveyed enterprises report that reasoning-class models materially accelerated their adoption~\cite{a16z2026}, and reasoning workloads consume substantially more output tokens per request---a fact that resurfaces in Section~\ref{sec:p2}.

\textbf{Price, long context, and multimodality.} Google's Gemini line competes on price-performance and context length (with long-context surcharges above 200K tokens), and on native multimodal processing~\cite{benchlm2026}.

\textbf{General knowledge work and ecosystem breadth.} OpenAI reports that 92\% of Fortune~500 companies touch ChatGPT and that it surpassed one million business customers in late 2025 (company-claimed); Microsoft reports more than 20 million paid Microsoft~365 Copilot seats~\cite{msft2026q3}.

The clearest micro-level evidence comes from deployers who publish their reasoning. Harvey, a legal AI platform originally built exclusively on OpenAI, added Anthropic and Google models in 2025 after internal benchmarking on legal tasks found there was ``no longer a single best model'': seven models---three from labs other than OpenAI---beat Harvey's original system on its BigLaw Bench, with Gemini strongest at drafting and OpenAI's o3 strongest at pre-trial analysis~\cite{harvey2025multimodel}. Harvey now routes tasks across all three labs. What a single sophisticated deployer discovered by measurement, the broader market discovered by procurement.

\subsection{Structural, Not Transitional}
\label{sec:structural}

An alternative reading holds that multi-lab licensing is a temporary artifact of an immature market that will consolidate as models commoditize. Three lines of evidence cut against this.

First, \emph{switching is rare while portfolios expand}: only 11\% of teams changed providers in a year, yet the share of enterprises using three or more families rose from 68\% to 81\% in roughly the same window~\cite{menlo2025,a16z2026}. Substitution-driven consolidation would show the opposite signature.

Second, \emph{enterprises are architecting for permanence}. JPMorgan built its LLM Suite portal---serving over 200{,}000 employees---explicitly to move across external models, with its Chief Data \& Analytics Officer stating the plan is ``not to be beholden to any one model provider''~\cite{cnbc2024jpm}. Goldman Sachs' firmwide assistant is architecturally model-agnostic across OpenAI, Google, Anthropic, and Meta models, a design the firm characterizes as risk mitigation against single-vendor dependence. These are capital investments in multi-lab operation, not stopgaps.

Third, \emph{capability divergence is widening at the frontier}, as the specialization evidence of \S\ref{sec:differentiation} shows, and bundling pressure from Microsoft and Google---which does exist---operates at the productivity-suite layer more than at the model-API layer.

\begin{figure}[t]
  \centering
  \includegraphics[width=\textwidth]{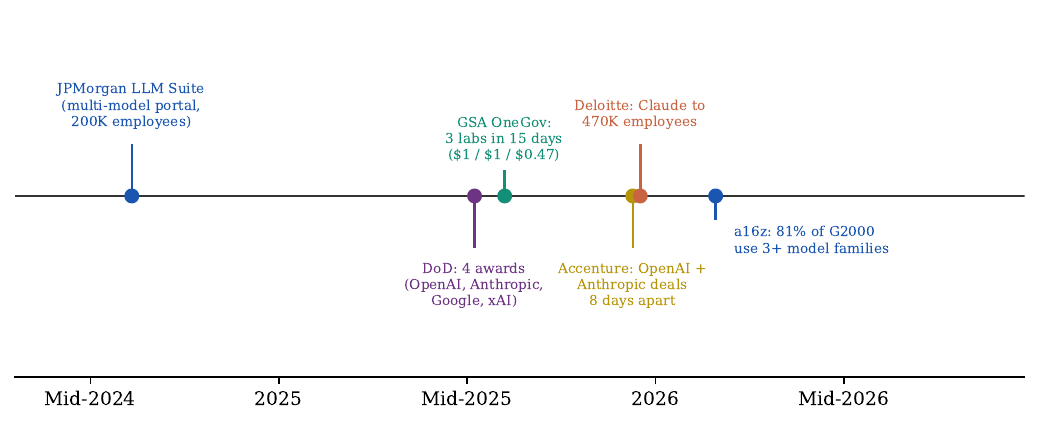}
  \caption{Selected multi-lab adoption events, 2024--2026. The clustering of multi-vendor commitments---three GSA agreements in fifteen days, two Accenture lab partnerships in eight days, four simultaneous DoD awards---indicates deliberate portfolio construction rather than sequential vendor evaluation. Sources: \cite{cnbc2024jpm,gsa2025onegov,accenture2025anthropic,a16z2026}.}
  \label{fig:timeline}
\end{figure}

The event record makes the pattern legible at a glance (Figure~\ref{fig:timeline}): across finance, consulting, and government, the signature is the same---multiple labs adopted nearly simultaneously, by design.

We therefore treat the multi-lab enterprise as the steady state for the foreseeable horizon, which makes the operational problems it creates worth solving rather than waiting out.

\subsection{The Phenomenon Across Verticals}
\label{sec:vertical}

Table~\ref{tab:verticals} summarizes documented multi-lab deployments across eight verticals. Three patterns are worth drawing out. In \emph{financial services}, the poles are visible: JPMorgan and Goldman Sachs institutionalized multi-lab architectures, while Morgan Stanley pursued a deliberate deep single-vendor partnership with OpenAI---the exception that proves the default. In \emph{consulting}, the multi-lab pattern is doubled: firms license multiple labs internally \emph{and} maintain delivery partnerships with all of them (Accenture announced OpenAI and Anthropic expansions within eight days of each other in December 2025~\cite{accenture2025anthropic}). In \emph{government}, the GSA OneGov agreements made the multi-lab default explicit policy, licensing all three labs for the entire federal workforce within a single month~\cite{gsa2025onegov}.

\begin{table}[t]
  \centering
  \caption{Documented multi-lab enterprise deployments by vertical (selected). Deployment modes: API = direct provider API; cloud = via Bedrock/Vertex/Azure; seats = SaaS subscriptions (ChatGPT Enterprise, Claude Enterprise, Gemini for Workspace).}
  \label{tab:verticals}
  \small
  \begin{tabularx}{\textwidth}{L{1.7cm} L{2.3cm} L{3.6cm} L{1.8cm} X}
    \toprule
    \textbf{Vertical} & \textbf{Organization} & \textbf{Providers} & \textbf{Mode} & \textbf{Stated rationale} \\
    \midrule
    Finance & JPMorgan (LLM Suite) & OpenAI first; multi-model by design & portal/API & ``Not beholden to any one model provider''~\cite{cnbc2024jpm} \\
    Finance & Goldman Sachs & OpenAI, Google, Anthropic, Meta & cloud & Model-agnosticism as risk mitigation \\
    Finance & Morgan Stanley & OpenAI & API/seats & Deliberate single-vendor depth (contrast case) \\
    Legal & Harvey & OpenAI, Anthropic, Google & cloud & Benchmarked task-level superiority per lab~\cite{harvey2025multimodel} \\
    Consulting & Accenture & OpenAI + Anthropic (+ Google partnership) & seats/API & Client-driven multi-partner strategy~\cite{accenture2025anthropic} \\
    Consulting & Deloitte & Anthropic (470K employees) + others & seats & Enterprise-wide rollout within multi-partner portfolio \\
    Healthcare & Mayo Clinic & Microsoft (Copilot, frontier partnership), Google Cloud, Abridge & seats/cloud & Different tools per clinical workflow \\
    Life sciences & Moderna & OpenAI (4{,}000+ custom GPTs) & seats & Scaling knowledge work without headcount growth \\
    Life sciences & Sanofi & OpenAI partnership + AWS & API/cloud & Drug-development acceleration \\
    Retail & Walmart & In-house model (``Wallaby'') & self-hosted & Build-vs-license contrast case~\cite{menlo2025} \\
    Government & U.S.\ federal (GSA OneGov) & OpenAI (\$1), Anthropic (\$1), Google (\$0.47) & seats & All-of-government access; agencies adopt any or all~\cite{gsa2025onegov} \\
    Defense & U.S.\ DoD & Anthropic, Google, OpenAI, xAI & contracts & Deliberately multi-vendor awards up to \$200M each \\
    \bottomrule
  \end{tabularx}
\end{table}

\section{P1: Governance Fragmentation}
\label{sec:p1}

Each additional licensed lab multiplies, rather than adds to, the governance burden. A three-lab enterprise must (i) perform three vendor security reviews against three distinct certification postures; (ii) reconcile three formats of model documentation---model cards and system cards are not standardized across labs in content, granularity, or update cadence; (iii) negotiate and monitor three data-processing agreements with different retention, training-use, and residency terms; and (iv) map all of this against the regulatory stack of Table~\ref{tab:frameworks}, whose obligations phase in on different clocks.

\begin{table}[t]
  \centering
  \caption{The overlapping governance framework stack facing a multi-lab enterprise.}
  \label{tab:frameworks}
  \small
  \begin{tabularx}{\textwidth}{L{3.1cm} L{2.4cm} X}
    \toprule
    \textbf{Framework} & \textbf{Status} & \textbf{Multi-lab implication} \\
    \midrule
    NIST AI RMF 1.0~\cite{nistairmf} & Voluntary (2023); de facto required for U.S.\ federal contractors & Govern/Map/Measure/Manage functions must be executed per model and per vendor \\
    ISO/IEC 42001:2023~\cite{iso42001} & Certifiable; increasingly a procurement requirement & AI management system must encompass every vendor in scope; audits multiply with vendor count \\
    EU AI Act~\cite{euaiact} & Mandatory; phased Feb 2025 -- Aug 2027 & Deployer obligations attach per system; shadow use of an unapproved lab can make the firm an unwitting deployer \\
    Sector: HIPAA / FDA & Mandatory (U.S.\ healthcare, life sciences) & BAAs and validation evidence required per vendor \\
    Sector: FINRA / SEC / OCC & Mandatory (U.S.\ financial services) & Model risk management (e.g., SR 11-7-style review) per external model family \\
    \bottomrule
  \end{tabularx}
\end{table}

\paragraph{Shadow AI as the multi-homing externality.} The same demand-side pull that drives sanctioned multi-lab licensing also drives unsanctioned use. Gartner projects that by 2030 more than 40\% of enterprises will experience a security or compliance incident linked to unauthorized shadow AI, and reports that 69\% of organizations already suspect or have evidence of prohibited public GenAI use~\cite{gartnershadow2025}. IBM's 2025 breach study found only 17\% of organizations have technical controls preventing employees from uploading confidential data to public AI tools, and 86\% lack visibility into AI data flows~\cite{ibmbreach2025}. Transaction data makes the financial face of this visible: one documented case surfaced \$120{,}000 in annual AI spend that appeared on no provider dashboard because it was purchased entirely on employee cards~\cite{rampblind2026}. Governance regimes designed for one sanctioned vendor systematically undercount an environment where employees, teams, and business units each add labs independently.

\paragraph{Why tooling alone has not solved P1.} A market of AI governance platforms (e.g., Credo AI, Holistic AI, IBM watsonx.governance, ModelOp) has emerged to manage inventories, policies, and attestations. These tools help, but they inherit the fragmentation they sit atop: they consume vendor-heterogeneous documentation and telemetry (P3), and they cannot normalize what providers do not expose. Governance fragmentation is thus not merely a policy problem; it is downstream of a data problem, a point we develop in Section~\ref{sec:discussion}.

\section{P2: FinOps Breakdown --- Balancing Tokens and Dollars}
\label{sec:p2}

\subsection{Why Tokens Defeat Budgets}

Traditional software budgeting assumes cost scales with a slow-moving driver: seats, servers, or headcount. Token-based AI consumption scales with \emph{behavior}. Agentic workflows consume roughly 7$\times$ the tokens of single-prompt usage; a tokenizer revision can change token counts for identical text by double-digit percentages; reasoning models multiply output tokens per request; and a single prompt-template change can reprice a production workload overnight~\cite{rampblind2026}. Across one large transaction panel, average monthly AI token spend grew 13$\times$ over eighteen months~\cite{ramp2026}.

\begin{figure}[t]
  \centering
  \includegraphics[width=0.82\textwidth]{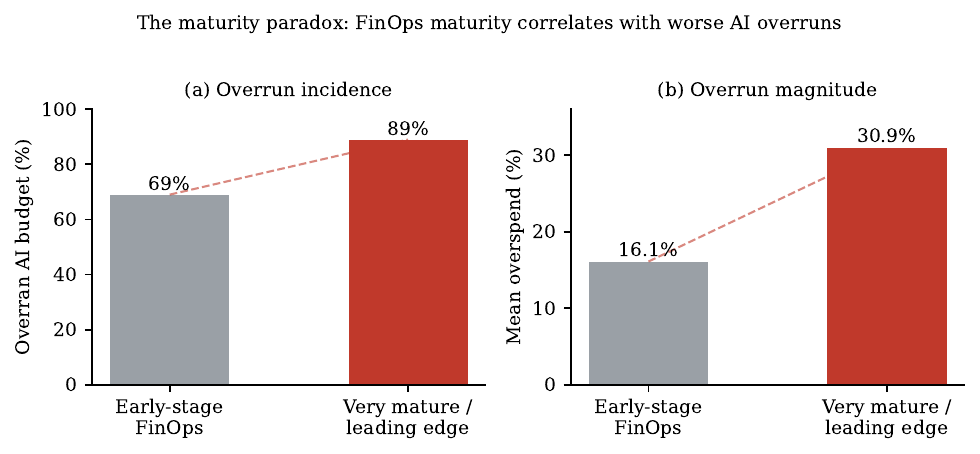}
  \caption{The maturity paradox in AI budgeting. Organizations self-assessing as FinOps-mature overran AI budgets more often (a) and by larger margins (b) than early-stage organizations. Survey of 500 enterprise finance leaders, $\pm$4.4pp at 95\% CI~\cite{doit2026}. A plausible mechanism: mature organizations deploy more AI, and practices tuned to provisioning-driven infrastructure costs transfer poorly to behavior-driven token costs.}
  \label{fig:overruns}
\end{figure}

The result is a budgeting failure documented across independent surveys. In a survey of 500 enterprise finance leaders, 79\% of organizations overran their AI budgets in the past year; strikingly, organizations self-assessing as ``very mature'' or ``leading edge'' in FinOps overran at a higher rate (89\%) and by a larger mean margin (30.9\%) than FinOps novices (69\%; 16.1\%)~\cite{doit2026}. We call this the \emph{maturity paradox} (Figure~\ref{fig:overruns}): existing FinOps practice, tuned to infrastructure whose costs scale with provisioning, transfers poorly to costs that scale with model behavior---and mature organizations deploy more AI, compounding exposure. A separate survey of 372 enterprises found 80\% missing AI infrastructure forecasts by more than 25\%, and half of companies with AI-core products not tracking LLM API costs at all~\cite{mavvrik2025}.

The canonical public case is Uber, whose CTO stated in May 2026 that the company had exhausted its full-year 2026 AI budget by April---``the budget I thought I would need is blown away already''---after rolling out agentic coding tools to roughly 5{,}000 engineers, with power users consuming \$500--\$2{,}000 per engineer per month; Uber responded by capping agentic-coding spend at \$1{,}500 per employee per month~\cite{forbes2026uber}.

\subsection{Cross-Provider Pricing Heterogeneity}

\begin{figure}[t]
  \centering
  \includegraphics[width=0.85\textwidth]{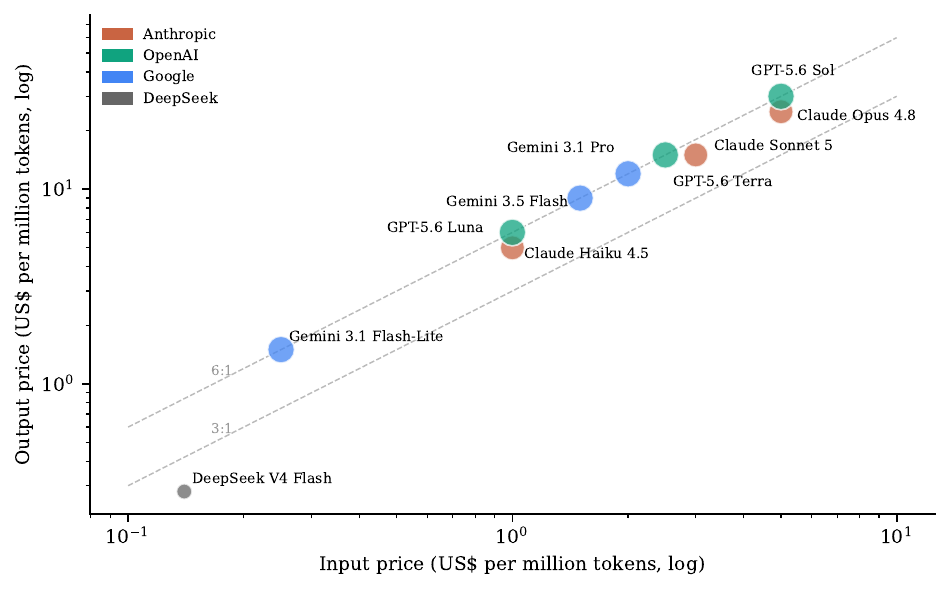}
  \caption{List API pricing across provider tiers, July 2026 (log--log). Bubble area scales with the output:input price ratio; dashed guides mark 3:1 and 6:1 ratios. Because reasoning and agentic workloads shift consumption toward output tokens, the ratio---not the headline input price---drives realized cost, and it varies across providers and tiers. Prices from third-party trackers synchronized to provider documentation~\cite{benchlm2026}; excludes caching, batch, long-context, and committed-use modifiers. Snapshot values; prices change frequently.}
  \label{fig:pricing}
\end{figure}

The multi-lab enterprise must reconcile not one but several pricing systems that differ on every axis: unit prices, discount mechanisms, and even the commercial model itself. Figure~\ref{fig:pricing} maps list per-token prices across the three labs' current model tiers: input prices span more than an order of magnitude, output prices nearly two, and---most consequentially for forecasting---the output:input ratio itself varies from 2:1 to 6:1 across tiers, so the same workload shifted between models changes not just its price level but its price \emph{structure}. Beneath list prices sit heterogeneous modifiers: prompt caching (discounts up to 90\% on cached input), batch processing (typically 50\%), long-context surcharges (Gemini~3.1~Pro doubles above 200K tokens), and provisioned-throughput commitments on each cloud (Azure PTUs, Bedrock provisioned throughput, GCP committed-use discounts)~\cite{benchlm2026}.

The commercial models diverge more fundamentally: ChatGPT Enterprise is seat-based with flat-rate usage; Claude Enterprise combines seats with a consumption-billed token pool; Google folded Gemini into Workspace seat pricing in 2025; Microsoft~365 Copilot is per-seat. Seat-dollars and token-dollars do not share a unit, which is precisely the ``balance'' problem P2 names: enterprises cannot compare, attribute, or charge back spend expressed in incommensurable units. The Tokenomics Foundation's extension of FOCUS to tokens~\cite{tokenomics2026} is the first credible standardization response, but it launched in 2026 and is unproven; notably, its initial supporters include both the largest buyers and the largest sellers of the problem.

\begin{figure}[t]
  \centering
  \includegraphics[width=0.85\textwidth]{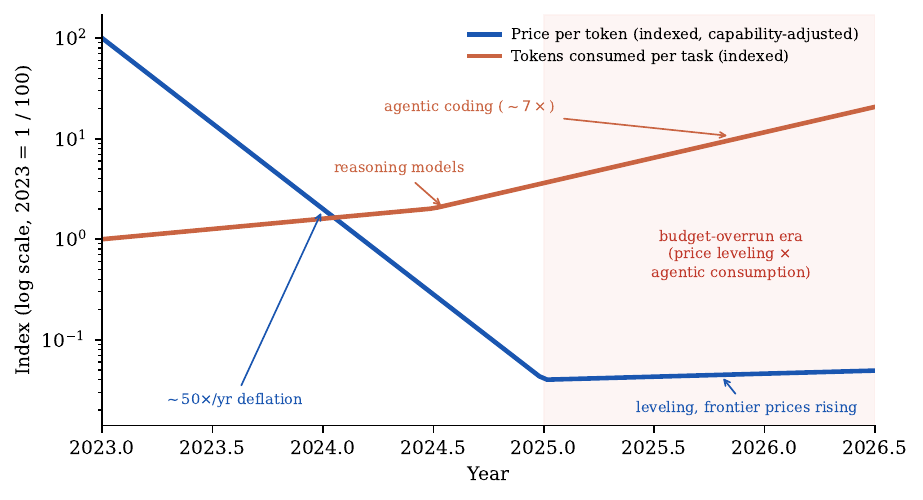}
  \caption{The token-economics scissors (stylized; indexed, log scale). Per-token prices fell $\sim$50$\times$/year through 2024~\cite{stanford2025index} then leveled, with frontier prices rising~\cite{finops2026}; tokens consumed per task rose with reasoning models and then sharply with agentic workflows ($\sim$7$\times$)~\cite{rampblind2026}. The shaded region marks the documented budget-overrun era (\S\ref{sec:p2}). Curves are illustrative reconstructions of cited trend magnitudes, not fitted series.}
  \label{fig:scissors}
\end{figure}

Historical context sharpens the problem rather than dissolving it. Inference prices for equivalent capability fell at a median of roughly 50$\times$ per year through 2024~\cite{stanford2025index}, but per-token costs leveled in 2025 and frontier-tier prices have since risen~\cite{finops2026}---while consumption per task grows with reasoning and agentic patterns (roughly 7$\times$ for agentic workloads). Figure~\ref{fig:scissors} depicts the resulting scissors: enterprises that budgeted on the deflation curve met the consumption curve, and the region where the curves diverge is precisely the 2025--2026 budget-overrun era documented above.

\section{Evaluating API and Agent Cost Burn}
\label{sec:agenteval}

Sections~\ref{sec:p2} and~\ref{sec:p3} established that agentic workloads are the fastest-growing driver of token burn and the hardest to see. This section addresses the layer in between: how an enterprise should \emph{evaluate} API usage and AI agents in economic terms---credits, tokens, and dollars---before and after deployment. Standard capability benchmarks are silent on this: a leaderboard rank says nothing about what a passing solution costs, and procurement decisions made on accuracy alone systematically select for expensive configurations.

\subsection{From Capability Benchmarks to Economic Evaluation}

A small but fast-growing literature has begun to correct this. Kapoor et al.\ argue that agent benchmarks which ignore cost invite artificially expensive leaderboard climbing (retries, ensembles, longer reasoning) and call for jointly reporting accuracy and dollar cost on a Pareto frontier~\cite{kapoor2024agents}. Erol et al.\ formalize the economic view with the \emph{cost-of-pass} metric: the expected dollar cost of obtaining a correct solution, defined as the cost of one attempt divided by the success rate, and the \emph{frontier cost-of-pass}: the minimum cost-of-pass achievable across available models, benchmarked against the cost of a human expert performing the same task~\cite{erol2025costofpass}. Applying this lens to agents on the GAIA benchmark, the OPPO AI Agent Team showed that component-level choices (backbone, planning depth, memory design, tool configuration) swing the economics far more than the accuracy: their cost-optimized framework retained 96.7\% of a leading open-source agent's performance while improving cost-of-pass by 28.4\% (\$0.398 to \$0.285 per task)~\cite{efficientagents2025}.

\begin{figure}[t]
  \centering
  \includegraphics[width=\textwidth]{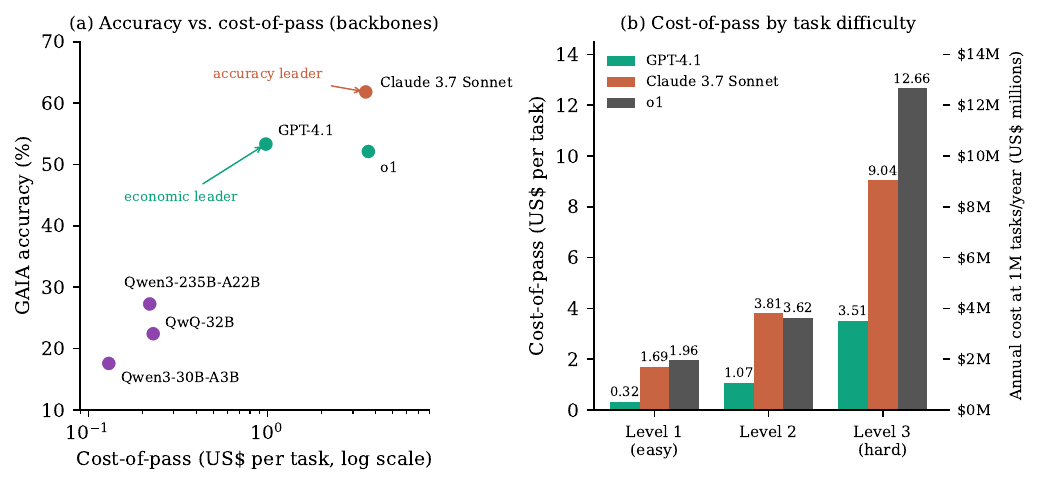}
  \caption{Agent cost economics on the GAIA benchmark; data from the Efficient Agents study~\cite{efficientagents2025} (per-token prices as of May 2025). (a)~Accuracy versus cost-of-pass for agent backbones: the accuracy leader (Claude~3.7 Sonnet) and the economic leader (GPT-4.1) are different models, and small open-weight models occupy a low-cost/low-accuracy corner. (b)~Cost-of-pass by GAIA difficulty level: economics deteriorate by 5--6$\times$ from easiest to hardest tier for frontier reasoning backbones; the right-hand axis translates per-task cost-of-pass into annual cost at a nominal one million tasks per year, expressed in US\$ millions, to convey enterprise-scale orders of magnitude. Model generations shown are those evaluated in the cited study; the pattern, not the point estimates, is the finding.}
  \label{fig:agentecon}
\end{figure}

Two of their empirical findings matter directly for the multi-lab enterprise. First, \emph{accuracy rankings and economic rankings disagree}: in their backbone comparison, Claude~3.7 Sonnet achieved the highest GAIA accuracy (61.8\% vs.\ GPT-4.1's 53.3\%) but at 3.6$\times$ the cost-of-pass (\$3.54 vs.\ \$0.98)---the ``best'' model was not the best buy for that workload~\cite{efficientagents2025}. Second, \emph{agent economics deteriorate nonlinearly with task difficulty}: from the easiest to the hardest GAIA tier, cost-of-pass rose 534\% for Claude~3.7 Sonnet and 646\% for OpenAI o1~\cite{efficientagents2025}. Figure~\ref{fig:agentecon} plots both findings. For a multi-lab enterprise this is the economic case for the portfolio itself---different labs occupy different points on the cost-accuracy frontier for different task classes---and simultaneously the case for measuring, because the frontier cannot be exploited blind.

\subsection{A Minimal Evaluation Framework for the Multi-Lab Enterprise}

Synthesizing this literature with the FinOps evidence of Section~\ref{sec:p2}, we propose that enterprises evaluate every API workload and agent deployment on five metrics (Table~\ref{tab:metrics}), computed per task class and per provider:

\begin{enumerate}
  \item \textbf{Cost per attempt} $C = n_{\mathrm{in}} \cdot p_{\mathrm{in}} + n_{\mathrm{out}} \cdot p_{\mathrm{out}} + c_{\mathrm{tools}}$, where $n$ are token counts, $p$ per-token prices, and $c_{\mathrm{tools}}$ tool/retrieval costs.
  \item \textbf{Cost-of-pass} $v = C / s$, the expected cost of a \emph{successful} outcome given success rate $s$~\cite{erol2025costofpass}. This is the number that belongs in a business case; $C$ alone is not.
  \item \textbf{Burn multiplier} $B = T_{\mathrm{agent}} / T_{\mathrm{single}}$, the ratio of tokens consumed by the agentic configuration to a single-shot baseline on the same task (empirically around 7$\times$ for agentic coding~\cite{rampblind2026}).
  \item \textbf{Difficulty elasticity}: the ratio of cost-of-pass on the hardest to the easiest tier of the workload (5--6$\times$ on GAIA for frontier backbones~\cite{efficientagents2025}); high elasticity means average-based budgets will be dominated by the tail.
  \item \textbf{Marginal cost of capability}: when comparing two models, the additional cost-of-pass per point of accuracy gained---making explicit the trade Kapoor et al.'s Pareto framing implies~\cite{kapoor2024agents}.
\end{enumerate}

\begin{table}[t]
  \centering
  \caption{A minimal cost-burn evaluation framework for enterprise API and agent workloads. All metrics are computed per task class and per provider, at current contract prices.}
  \label{tab:metrics}
  \small
  \begin{tabularx}{\textwidth}{L{3.2cm} L{4.6cm} X}
    \toprule
    \textbf{Metric} & \textbf{Definition} & \textbf{Decision it informs} \\
    \midrule
    Cost per attempt ($C$) & $n_{\mathrm{in}} p_{\mathrm{in}} + n_{\mathrm{out}} p_{\mathrm{out}} + c_{\mathrm{tools}}$ & Rate-limit and budget-cap settings; anomaly thresholds \\
    Cost-of-pass ($v$) & $C / s$ for success rate $s$~ & Model selection per task class; build-vs-buy; ROI baselines vs.\ human-expert cost \\
    Burn multiplier ($B$) & Agent tokens $\div$ single-shot tokens on the same task & Whether agentic autonomy is worth its overhead for this task class \\
    Difficulty elasticity & $v_{\mathrm{hard}} / v_{\mathrm{easy}}$ across workload tiers & Forecast realism; whether to cap or route hard-tier tasks \\
    Marginal cost of capability & $\Delta v / \Delta$accuracy between candidate models & When to pay for the frontier model vs.\ a mid-tier one \\
    \bottomrule
  \end{tabularx}
\end{table}

\subsection{A Worked Example}
\label{sec:worked}

To make the framework concrete, consider an enterprise evaluating three configurations for an internal ticket-resolution agent, using July 2026 list prices (Figure~7) and \emph{illustrative but realistic} consumption and success parameters; the arithmetic, not the parameters, is the point.

\begin{itemize}
  \item \textbf{Config A (frontier mid-tier):} Claude Sonnet~5 (\$3 in / \$15 out per M tokens); 40K input + 8K output tokens per attempt $\Rightarrow C_A = \$0.24$; success $s_A = 0.75$ $\Rightarrow v_A = \$0.32$.
  \item \textbf{Config B (efficient tier):} GPT-5.6 Luna (\$1 / \$6); weaker model iterates more: 60K + 12K tokens $\Rightarrow C_B = \$0.132$; $s_B = 0.45$ $\Rightarrow v_B = \$0.293$.
  \item \textbf{Config C (budget tier):} Gemini 3.1 Flash-Lite (\$0.25 / \$1.50); 80K + 16K tokens $\Rightarrow C_C = \$0.044$; $s_C = 0.12$ $\Rightarrow v_C = \$0.367$.
\end{itemize}

Three lessons fall out. First, the \emph{rank reversal}: Config~C is 5.5$\times$ cheaper than Config~A per attempt yet the most expensive per resolved ticket---exactly the inversion the GAIA data shows at benchmark scale. Second, the \emph{cascade dominates}: routing every ticket to Config~B first and escalating failures to Config~A yields expected cost \$0.264 and success $0.45 + 0.55 \times 0.75 = 0.86$, i.e.\ $v = \$0.306$---near Config~B's economics at well above Config~A's success rate, which is the enterprise-scale version of the routing results in Section~\ref{sec:background} and a direct economic argument for the multi-lab portfolio. Third, the evaluation is impossible without P3: computing $s$ and $n$ per task class requires exactly the per-request, per-outcome telemetry that fragmented provider surfaces do not natively supply---which is why we place this section between the FinOps problem and the telemetry problem.

\section{P3: Telemetry Fragmentation - Measuring Adoption and Spend Across Labs}
\label{sec:p3}

If P2 is about incommensurable dollars, P3 is about incommensurable data. Table~\ref{tab:telemetry} compares the administrative telemetry surfaces of the three leading labs as of July 2026.

\begin{table}[t]
  \centering
  \caption{Provider administrative telemetry surfaces (July 2026). Compiled from provider documentation~\cite{anthropicusagecost,openaicompliance} and third-party analyses; details change frequently.}
  \label{tab:telemetry}
  \small
  \begin{tabularx}{\textwidth}{L{2.6cm} X X X}
    \toprule
    & \textbf{OpenAI (ChatGPT Ent.)} & \textbf{Anthropic (Claude)} & \textbf{Google (Gemini)} \\
    \midrule
    Usage analytics & Admin console dashboard; on-demand CSV exports (users, GPTs, projects) & Analytics API (aggregated usage/cost; separate key type from Admin API) & Workspace admin reporting for Gemini; Cloud billing for API \\
    Audit/compliance & Compliance API + admin audit logs & Separate Compliance API (per-event records) & Cloud audit logging \\
    Cost granularity & Seat-based; no per-token end-user charges & Per-user and per-SCIM-group cost; org/user spend limits with threshold alerts & Seat-bundled (Workspace) + per-token (Cloud API) \\
    Refresh latency & Near-real-time dashboard & 4--24h refresh; figures revisable for up to 30 days & Billing-cycle latency on Cloud side \\
    Known blind spots & Consumption invisible in dollar terms (flat seats) & Claude Code via Bedrock absent from Analytics API (must use AWS cost data) & Split across Workspace vs.\ Cloud surfaces \\
    ``Active user'' definition & Provider-defined & Provider-defined (differs) & Provider-defined (differs) \\
    \bottomrule
  \end{tabularx}
\end{table}

Three properties of this fragmentation deserve emphasis. First, the surfaces differ not only in format but in \emph{ontology}: seat-based products report users and conversations, API products report tokens and requests, and no two providers define an ``active user'' identically, so cross-lab adoption comparisons are category errors by construction. Second, the surfaces differ in \emph{temporal semantics}: one lab's near-real-time dashboard cannot be joined cleanly with another's figures that refresh on a 4--24 hour cycle and remain revisable for 30 days~\cite{anthropicusagecost}. Third, routing paths create \emph{blind spots}: consumption of the same model through a cloud marketplace can vanish from the lab's own analytics surface entirely, reappearing only in the cloud provider's billing data.

Enterprises respond by building unification layers from a rapidly growing third-party stack: LLM gateways that normalize requests and capture spend at the point of routing (LiteLLM, OpenRouter, Portkey, Kong AI Gateway, Cloudflare AI Gateway); observability platforms that trace usage and cost per feature (Langfuse, LangSmith, Helicone, Datadog LLM Observability, Arize, Weights \& Biases Weave); and cost platforms extending cloud FinOps to AI (CloudZero, Vantage, Finout). The gateway pattern is powerful precisely because it restores a single point of measurement---but it covers only API traffic that the enterprise routes through it, leaving seat-based products (where much adoption lives) and shadow usage outside the frame.

The stakes of not measuring are documented in unit-economics failures: GitHub Copilot reportedly lost around \$20 per user per month in its early period against a \$10 subscription; a single agentic query can fan out into dozens of LLM calls invisible to the feature owner until the invoice arrives~\cite{rampblind2026}. The measurement gap, in other words, is not an accounting inconvenience; it determines whether AI products and internal deployments have knowable economics at all.

\section{Discussion: P1--P3 as One Missing Abstraction}
\label{sec:discussion}

The three problems are usually assigned to three different organizational owners---governance to risk and compliance, spend to finance, telemetry to platform engineering---and to three different tool markets. We argue they are facets of a single missing abstraction: a \emph{cross-provider enterprise AI control plane}, a normalized layer through which policy, spend, and usage for all licensed labs are expressed in shared units.

The argument is structural. P1 cannot be solved without P3: governance requires knowing what models are used, by whom, for what---exactly the data the fragmented telemetry surfaces fail to provide consistently, and shadow usage evades entirely. P2 cannot be solved without P3 either: cost attribution and chargeback presuppose usage data in commensurable units. And P3 alone is insufficient without the policy (P1) and financial (P2) semantics layered on top, which is why raw gateways and dashboards have not closed the gap. Every existing tool category solves a slice---gateways normalize API traffic, FOCUS/Tokenomics normalizes cost records, governance platforms normalize attestations---but nothing integrates the three, and each layer inherits the blind spots of the layers below.

\subsection{A Token-Spend Governance Framework}
\label{sec:tsg}

We propose a six-layer Token-Spend Governance (TSG) framework as the enterprise-side instantiation of the control plane (Figure~\ref{fig:cockpit}a). The framework deliberately composes standards and results already established in this paper rather than inventing new ones: its lower layers operationalize P3, its middle layers P2, and its upper layers P1.

\textbf{L0 --- Data.} Ingest every spend- and usage-bearing source: provider analytics and compliance APIs, cloud-marketplace billing (Bedrock, Vertex, Azure), LLM-gateway logs, and corporate-card transaction feeds. The card feed is not optional; it is the only source that captures shadow AI (\S\ref{sec:p1}).

\textbf{L1 --- Normalization.} Express all records in shared units via the FOCUS specification and its token extension~\cite{tokenomics2026}, supplemented by the three definitions Section~\ref{sec:discussion} argues have no current owner: a tokenizer-adjusted normalized token, an auditable active-user definition spanning seat- and API-based products, and a task-class taxonomy.

\textbf{L2 --- Measurement.} Compute the five cost-burn metrics of Table~\ref{tab:metrics} per task class and per provider, with cost-of-pass~\cite{erol2025costofpass} as the primary decision metric.

\textbf{L3 --- Allocation.} Attribute normalized spend to business units, features, and individual agents (chargeback or showback), and manage commitment vehicles (PTUs, provisioned throughput, committed-use discounts) against realized consumption.

\textbf{L4 --- Control.} Enforce budgets and caps in near-real time---organization, team, and per-user limits with threshold alerts; anomaly detection tuned to behavior-driven cost (a prompt-template change, an agent loop); and policy-enforced routing, including the cascade strategies of Section~\ref{sec:worked}. Uber's \$1{,}500/employee/month agentic cap~\cite{forbes2026uber} is an L4 control retrofitted after an L2 failure; TSG's claim is that the layers must exist \emph{before} deployment scales.

\textbf{L5 --- Assurance.} Maintain audit trails and map them to the governance stack of Table~\ref{tab:frameworks}: NIST AI RMF functions, ISO/IEC 42001 audit evidence, and EU AI Act deployer obligations, per vendor.

The framework carries a maturity dimension (M0--M4, Figure~\ref{fig:cockpit}a): from \emph{invisible} (no unified view---the documented state of half of AI-core companies~\cite{mavvrik2025}), through \emph{visible} (one normalized view), \emph{attributed} (chargeback live), and \emph{governed} (caps and alerts active), to \emph{optimized} (cost-of-pass-driven routing across the lab portfolio). The maturity paradox of Section~\ref{sec:p2} carries a warning for this ladder: cloud-FinOps maturity does not transfer automatically; each rung must be rebuilt on token-native units.

\subsection{A Unified FinOps--Governance Cockpit (Concept)}
\label{sec:cockpit}

\begin{figure}[t]
  \centering
  \includegraphics[width=\textwidth]{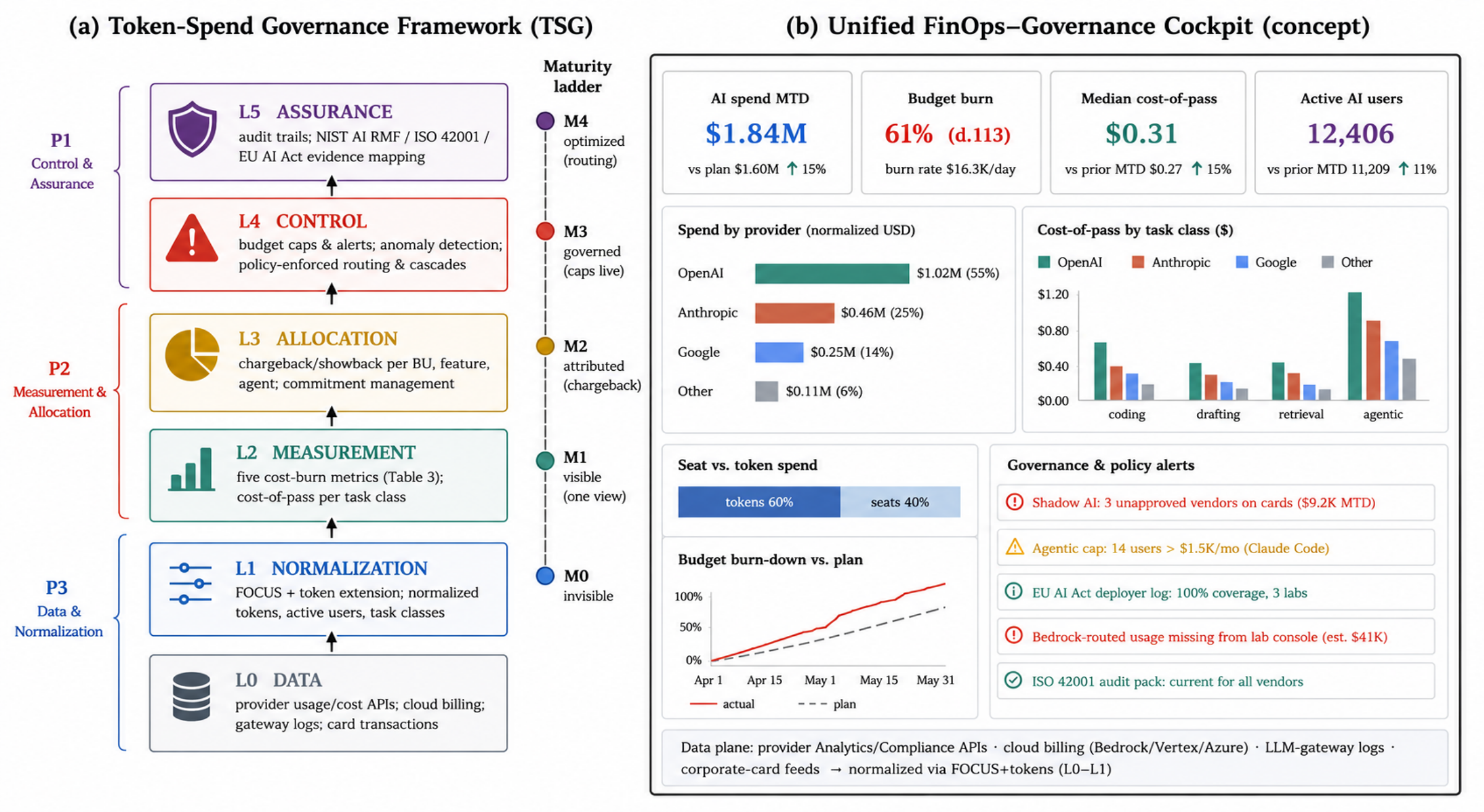}
  \caption{Conceptual proposal for enterprises (not a product). (a)~The six-layer Token-Spend Governance (TSG) framework with its maturity ladder (M0 invisible $\to$ M4 optimized); lower layers operationalize P3 (data and normalization), middle layers operationalize P2 (measurement and allocation), and upper layers operationalize P1 (control and assurance). (b)~Conceptual wireframe of a unified FinOps--Governance cockpit integrating all licensed frontier AI labs. The dashboard presents normalized headline economics, per-provider spend, seat-versus-token composition, budget burn-down, cost-of-pass by task class and provider, and governance alerts including shadow AI, agentic spending caps, cloud-routing blind spots, and compliance evidence. All numerical values are illustrative and shown for conceptual purposes only.}
  \label{fig:cockpit}
\end{figure}

Figure~\ref{fig:cockpit}b sketches, as a concept only, the visualization layer TSG implies: a single cockpit integrating all licensed frontier labs, oriented primarily toward FinOps governance rather than raw usage telemetry. Four design commitments distinguish it from existing per-vendor consoles and gateway dashboards. First, \emph{normalized headline economics}: total spend, budget burn versus plan, median cost-of-pass, and active users are computed in L1 units, so the numbers are comparable across seat-based and token-based products---something no provider console can offer for its competitors. Second, \emph{task-class economics per lab}: cost-of-pass by workload class and provider, which is simultaneously an optimization view (where routing would save money) and the empirical justification for the multi-lab portfolio itself (\S\ref{sec:differentiation}). Third, \emph{governance alerts as first-class citizens}: shadow-AI detections from the card feed, agentic-cap breaches, cloud-routed consumption missing from lab consoles (the Bedrock blind spot of Table~\ref{tab:telemetry}), and compliance-evidence status per framework---placing P1 on the same pane of glass as P2. Fourth, \emph{provenance-labeled data}: every number carries its source and refresh latency, because Section~\ref{sec:p3} showed the surfaces disagree on both. We emphasize that Figure~\ref{fig:cockpit} is a conceptual proposal for enterprises and a design target for the tooling ecosystem, not a product; its purpose in this paper is to make the control-plane abstraction concrete enough to critique, build, and standardize against.

From this diagnosis follows a concrete agenda:

\begin{enumerate}
  \item \textbf{A minimal cross-provider metric standard.} Three definitions would unlock disproportionate value: (i) a normalized token accounting for tokenizer differences (or a task-normalized consumption unit); (ii) a standard, auditable definition of an active AI user across seat- and API-based products; (iii) a shared task-class taxonomy (e.g., coding, drafting, retrieval, agentic execution) so that adoption and spend can be compared across labs by workload. The Tokenomics/FOCUS effort~\cite{tokenomics2026} addresses (i) for cost records; (ii) and (iii) have no current owner.
  \item \textbf{Provider telemetry parity.} Regulatory deployer obligations (e.g., under the EU AI Act) implicitly require usage visibility that current provider surfaces do not uniformly supply; procurement leverage---especially from multi-lab buyers like the GSA---could make telemetry completeness (including cloud-routed consumption) a standard contractual term.
  \item \textbf{Research.} The routing literature should be extended upward: from cost-quality routing per query to governance- and budget-constrained routing per organization, evaluated not only on benchmark quality but on auditability, attribution accuracy, and forecast error. Empirically, longitudinal firm-level studies of multi-lab portfolios---entry, expansion, and the rare exits---would put the ``structural, not transitional'' claim of \S\ref{sec:structural} on firmer causal footing.
\end{enumerate}

\section{Conclusion}
\label{sec:conclusion}

The enterprise AI market has settled, for now, into a configuration few procurement playbooks anticipated: the typical large enterprise licenses every top-tier frontier lab because the labs are genuinely good at different things, and the buyers know it. This paper documented the phenomenon across survey, transaction, and disclosure data and across eight verticals, and characterized the three operational problems it creates---governance fragmentation, FinOps breakdown, and telemetry fragmentation. Each problem is real on its own terms: multiplied compliance surfaces and proliferating shadow use; near-universal budget overruns that worsen, paradoxically, with FinOps maturity; and adoption data that cannot be compared across the very vendors being adopted. But the deeper finding is their unity. An enterprise that cannot measure consistently cannot govern or budget consistently. Until a cross-provider control plane---standards, telemetry parity, and integrated tooling---emerges, every multi-lab enterprise will keep rebuilding, privately and partially, the same missing layer; the Token-Spend Governance framework and cockpit concept of Sections~\ref{sec:tsg}--\ref{sec:cockpit} are offered as a starting design for that layer. The market has voted for many models; the operating model for many models does not yet exist. Building it is a tractable, consequential agenda for researchers, standards bodies, and the labs themselves.

\paragraph{Data availability.} All quantitative claims cite public sources; no proprietary data was used.

\bibliographystyle{unsrtnat}
\bibliography{references}

\end{document}